# Investigation and field effect tuning of thermoelectric properties of SnSe$_2$ flakes

Ilaria Pallecchi, Federico Caglieris, Michele Ceccardi, Nicola Manca, Daniele Marré, Luca Repetto, Marine Schott
*CNR-SPIN and University of Genoa, Via Dodecaneso 33, 16146, Genoa, Italy*

Daniel I. Bilc
*Faculty of Physics, Babeş-Bolyai University, 1 Kogălniceanu, RO-400084 Cluj-Napoca, Romania*
*Independent Contribution, RO-407280 Florești, Romania*

Stefanos Chaitoglou, Athanasios Dimoulas
*Institute of Nanoscience and Nanotechnology, National Center for Scientific Research 'DEMOKRITOS', 15310, Athens, Greece*

Matthieu J. Verstraete
*Nanomat, Q-Mat, CESAM, and European Theoretical Spectroscopy Facility, Université de Liège (B5), B-4000 Liège, Belgium*

**Abstract**
The family of Van der Waals dichalcogenides (VdWDs) includes a large number of compositions and phases, exhibiting varied properties and functionalities. They have opened up a novel electronics of two-dimensional materials, characterized by higher integration and interfaces which are atomically sharper and cleaner than conventional electronics. Among these functionalities, some VdWDs possess remarkable thermoelectric properties. SnSe$_2$ has been identified as a promising thermoelectric material on the basis of its estimated electronic and transport properties. In this work we carry out experimental measurements of the electric and thermoelectric properties of SnSe$_2$ flakes. For a 30 micron thick SnSe$_2$ flake at room temperature, we measure electron mobility of 40 cm$^2$ V$^{-1}$ s$^{-1}$, a carrier density of 4 x 10$^{18}$ cm$^{-3}$, a Seebeck coefficient $S \approx -400$ µV/K and thermoelectric power factor $S^2\sigma \approx 0.35$ mW m$^{-1}$ K$^{-2}$. The comparison of experimental results with theoretical calculations shows fair agreement and indicates that the dominant carrier scattering mechanisms are polar optical phonons at room temperature and ionized impurities below 50 K. In order to explore possible improvement of the thermoelectric properties, we carry out reversible electrostatic doping on a thinner flake, in a field effect setup. On this 75 nm thick SnSe$_2$ flake, we measure a field effect variation of the Seebeck coefficient of up to 290 % at low temperature, and a corresponding variation of the thermoelectric power factor of up to 1050 %. We find that the power factor increases with the depletion of n-type charge carriers. Field effect control of thermoelectric transport opens perspectives for boosting energy harvesting and novel switching technologies based on two-dimensional materials.

**Introduction**
Graphene and related two-dimensional compounds represent a unique material platform for multifold technological applications, ranging from electronics, optoelectronics or spintronics to energy generation, energy storage, and sensors. As a further functionality that has not yet been widely explored, Van der Waals dichalcogenides (VdWDs) present features which should make them good thermoelectrics [1]. Indeed, VdWDs have an electronic band gap in the visible spectrum, with both n-type and p-type transport, quite large effective masses, and band degeneracy. Additionally, they have moderate thermal conductivity (few 10s of W m$^{-1}$ K$^{-1}$), which can be further reduced by microstructuring or epitaxy engineering [2,3,4]. Their electronic band structures and phonon spectra are sensitive to confinement [5,6], so that their thermoelectric properties can be tuned by varying sample thickness in the range of a few atomic layers. The thermal conductivity κ was found to decrease with thickness in SnSe$_2$ [7], but, on the other hand, it was demonstrated that thickness must not necessarily be in the few atomic layers range to have the lowest κ. In MoSe$_2$ κ exhibits a saturating behavior with decreasing thickness below 30 atomic layers [8]. Extremely large tunability of transport properties is also obtained by field effect in flakes just a few atomic layers thick, which allows to identify optimal values of the chemical potential to boost thermoelectric performance. These field effect experiments show that values of thermoelectric power factors of VdWDs at room temperature can be comparable to those of commercial thermoelectrics [9,10,11].

In previous work, we carried out theoretical studies to identify those VdWDs that possess the highest thermoelectric performances. WSe$_2$, MoTe$_2$, SnSe$_2$ were predicted to exhibit remarkable values of thermoelectric power factor and figure of merit, further tunable by thickness and lattice strain [12]. Among these VdWDs, the thermoelectric properties of SnSe$_2$ appear to be particularly promising, and this compound is chemically akin to its layered monochalcogenide counterpart SnSe. The latter's thermoelectric figure of merit reaches 2.6 at 923 K in p-type single crystals, and was a breakthrough in the search for new thermoelectric materials [13,14]. Remarkably in view of applications, both p-type and n-type SnSe polycrystals can be optimized for high temperature operation by chemical doping [15,16]. Nanostructuring was also attempted to improve thermoelectric performance of SnSe films [17] and nanoflakes [18,19]. In its dichalchogenide sibling SnSe$_2$, thermoelectric properties of samples in single crystalline form were experimentally explored only by Pham et al. [20]. The high quality SnSe$_2$ flakes in their work exhibited thermoelectric power factors (product of squared Seebeck coefficient $S^2$ and electrical conductivity $\sigma$) $S^2\sigma \sim 0.15$ mW m$^{-1}$ K$^{-2}$ at room temperature and 0.34 mW m$^{-1}$ K$^{-2}$ at 673 K in the in-plane direction, while the corresponding figures of merit ZT= $S^2\sigma T/\kappa$ ($\kappa$ thermal conductivity, T absolute temperature) were ZT~0.01 and 0.1 at room temperature and 673 K, respectively. Their analysis of angle-resolved photoemission spectroscopy (ARPES) indicated that the thermoelectric performance might be further improved by electron doping. The effect of chemical doping was exclusively explored in samples fabricated in the form of highly oriented nanosheet pellets by Spark Plasma Sintering (SPS) [21,22,23]. Undoped, chlorine-doped, barium-doped and Cu-intercalated SnSe$_2$ pellets all resulted to be n-type. Enhanced power factors along the in-plane direction were achieved by simultaneously introducing Se deficiency and chlorine doping, while reduced thermal conductivity was obtained thanks to efficient phonon scattering at grain boundaries. At room temperature, $S^2\sigma \approx 0.8$ mW m$^{-1}$K$^{-2}$ and $\kappa \approx 2$ Wm$^{-1}$K$^{-1}$ resulted in ZT$\approx$0.15, and at 673 K a remarkable ZT$\approx$0.63 was obtained [22]. Simultaneous copper intercalation and barium substitution led to a power factor $\approx$1.2 mW m$^{-1}$K$^{-2}$, almost temperature-independent from 300 to 773 K, which was the highest reported for all polycrystalline SnSe$_2$- and SnSe-based materials, with corresponding figures of merit ZT $\approx$0.10 at 300 K and ZT $\approx$0.67 at 773 K [23]. The effect of the microstructure on thermoelectric properties was also investigated by Chen et al. in SnSe$_2$ films deposited by physical vapor deposition and annealing in Se partial pressure [24]. In these films, structural disorder was introduced by random in-plane orientation among successive layers, while maintaining a high degree of c-axis orientation. After Se annealing, which mitigated Se vacancies and promoted the SnSe-to-SnSe$_2$ phase transition, cross-plane Seebeck coefficients up to -630 $\mu$V/K and power factors up to 2x10$^{-4}$ mW m$^{-1}$K$^{-2}$ at room temperature were measured.

From the above summary of available literature on theoretical and experimental studies on SnSe$_2$, it emerges that this compound has strong potential as a thermoelectric material, but the effect of doping on single crystalline SnSe$_2$ flakes has not been investigated. Since SnSe$_2$ and other high potential dichalcogenides (WSe$_2$, MoTe$_2$) have low dielectric permittivities, their polycrystalline forms will have significant grain boundary electrical resistance due to imperfect screening at interfaces, which will decrease their thermoelectric performance. For such materials, nanostructuring achieved by grain size minimization is not necessarily the best route to improve thermoelectric performance. For this reason, the optimization of single crystalline SnSe$_2$ flakes is very important [12,25]. Furthermore, a field effect setup offers the opportunity to carry out clean and reversible doping to tune electric and thermoelectric properties of SnSe$_2$ flakes by band filling. In order to implement this strategy and confirm theoretical predictions [12], in this work we carry out experimental measurements of the electric and thermoelectric properties of micrometric thick SnSe$_2$ flakes and explore possible improvement of these properties in thinner SnSe$_2$ flakes under field effect. We compare the experimental results with theoretical calculations including quantitative doping and scattering rates for SnSe$_2$.

**Experimental methods**

Crystals of 1T-SnSe$_2$ were purchased from HQ Graphene, and flakes of different thickness and lateral size were mechanically exfoliated and transferred for characterization onto SrTiO$_3$ substrates (see below). Larger flakes of micrometric thickness and hundreds of microns lateral size were mechanically exfoliated and transferred on quartz substrates. Additional capping layers of hexagonal boron nitride (h-BN) were transferred on top of these large flakes, as protection against charged surface impurities. For smaller flakes of few tens of nm thickness and few tens of micron lateral size, SrTiO$_3$ was chosen as a substrate suitable for field effect experiments, due to its high relative dielectric permittivity $\varepsilon_r$~300 at room temperature and up to $\varepsilon_r$~20000 at low temperature and low electric field [26]. Patterns of micrometric electrical contacts were realized by optical lithography, thermal evaporation of CrO/Au and lift off. CrO/Au serpentines were also fabricated and calibrated to be used as heaters and thermometers for thermoelectric measurements in flakes of micrometric size. The exfoliated flakes were transferred onto the micrometric patterns using a micromanipulator. Successively, finer electrical contacting was realized by Focused Ion Beam deposition of metallic tungsten. Flake thicknesses were measured from Atomic Force Microscope (AFM) profiles.

Resistance, Hall effect, magnetoresistance, and thermopower measurements from room temperature down to 10 K and in magnetic fields up to 9 T were carried out in a commercial PPMS by Quantum Design, with home-made adaptations to the sample holder and acquisition software. A four-probe geometry was used for electrical contacts. Thermopower measurements were performed in high vacuum and a steady-state technique was used to feed heat into the free-standing sample and establish a temperature gradient. A CrO/Au serpentine heater was placed on one sample edge, while the opposite edge was linked to the thermal sink through a clamp. A sinusoidal current with 100 s period was applied to the heater, and the response extracted by Fast Fourier Transform of the voltage signals measured at the leads and of the temperature signals. The latter were measured using the calibrated CrO/Au thermometers (for flakes of tens of micron size) or with thermocouples (for flakes of hundreds of micron size), respectively.

Electric and thermoelectric transport measurements were also carried out under field effect, by applying a voltage to a metallic gate electrode placed on the back of the 0.5 mm thick SrTiO$_3$ substrate. A positive (negative) gate electric field accumulates (depletes) n-type mobile charge carriers in the VdWD at the VdWD/SrTiO$_3$ interface, thus tuning the electric and thermoelectric properties, which depend on the carrier concentration. As the electric field decays into the VdWD, due to screening by mobile charge carriers, the relative tuning of the electric and thermoelectric properties depends on how deep the electric field penetrates into the VdWD as compared to the total thickness of the VdWD flake itself; hence a larger relative tuning is obtained for flakes of smaller thickness and smaller carrier density.

Figure 1 shows a contacted micrometric flake. Due to the irregular sample shape, we estimate a 30% uncertainty on the magnitude of electrical transport properties, related to geometrical factors. Furthermore, from the measurement of a sizeable number of SnSe$_2$ samples, we observe also some sample-to-sample variations in the exact temperature dependence of electric and thermoelectric transport properties, which could be due to accidental factors at the micrometric or nanometric scale, such as bending of the flake or inhomogeneity in the impurity distribution.

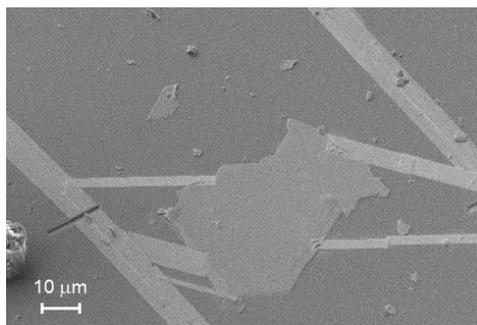

**Fig. 1**: SEM image of the exfoliated SnSe$_2$ flake of micrometric size within a pattern of contacts realized by optical lithography and FIB.

**Theoretical methods**

The electronic and thermoelectric properties of SnSe$_2$ bulk were studied within density functional theory (DFT) formalism using the B1-WC hybrid functional [27]. B1-WC describes the electronic (band gaps) and structural properties with better accuracy than the usual semi-local exchange-correlation functionals, and it is more appropriate for correlated materials with *d* electronic states [28,29]. The electronic structure calculations were performed using the linear combination of atomic orbitals method as implemented in the CRYSTAL code [30]. We use localized Gaussian-type basis sets including polarization orbitals and considered all the electrons for Se [31], and a Hartree-Fock pseudopotential for Sn [32]. The exponents of the most diffuse valence and polarization Gaussian functions were optimized within B1-WC to minimize the total energy of SnSe$_2$ bulk. The optimized exponents are: 1.5963, 0.7686, 0.3199, 0.1, 1.5136, 1.0497, and 0.2607 for 4s, 4p, 5sp, 6sp, 4d, 5d, and 6d functions of Se; 2.5335, 0.2056, 0.8050, and 0.2259 for 5p, 6sp, 5d, and 6d functions of Sn. To increase the efficiency of the CRYSTAL code, the two outermost s and p functions were combined into 5sp (for Se) and 6sp functions (for Se and Sn).

Brillouin zone integration was performed using a 6×6×6 mesh of k-points. The self-consistent-field calculations were considered to be converged when the energy changes between interactions were smaller than $10^{-8}$ Hartree. An extra-large predefined and pruned grid consisting of 75 radial points and 974 angular points was used for the numerical integration of the charge density. Full optimizations of the lattice constants and atomic positions were performed with the convergence criteria of 5×10$^{-5}$ Hartree/Bohr in the root-mean square values of forces and 1.2×10$^{-3}$ Bohr in the root-mean square values of atomic displacements. The level of accuracy in evaluating the Coulomb and exchange series is controlled by five tolerance parameters ($10^{-ITOLj}$, j=1-5). The ITOL values used in our calculations were 7, 7, 7, 9, and 30.

The transport properties were calculated within the Boltzmann transport formalism and the constant relaxation time approximation, using the BoltzTraP transport code [33]. We fixed the relaxation times on the basis of the experimental resistivity curves and included a generic *1/T* temperature dependence for the acoustic and optical phonon scatterings at high *T*. We use $\tau_{ac} = \tau_{ac}^{ref}(T/T_{ref})^{-1}E_r^{-1/2}$ for the degenerate case, and $\tau_{ac} = \tau_{ac}^{ref}(T/T_{ref})^{-3/2}E_r^{-1/2}$ for the nondegenerate case, then $\tau_{po} = \tau_{po}^{ref}(T/T_{ref})^{-1/2}E_r^{1/2}$ for the polar optical phonon scattering. Ionized impurity relaxation times are given by $\tau_{imp} = \tau_{imp}^{ref}(T/T_{ref})^{3/2}E_r^{3/2}$. In all of the above, $T_{ref}$ is a reference temperature at which the scattering times are known (calculated or fit to experiment), and $E_r$ is the reduced energy $E_r = E/k_BT$ [34,35,36]. The transport coefficients were very well converged using energies calculated on a mesh of 47×47×41 k-points.

## Results and discussion

*a) Bulk SnSe$_2$ flake*

We first present the electric and thermoelectric characterization of a large SnSe$_2$ flake of 30 μm thickness and hundreds of μm lateral size, representative of bulk-like behavior, capped with h-BN to preserve high mobility. Fig. 2 (main panel) shows resistivity measurements of the SnSe$_2$/BN flake, exhibiting a semiconducting temperature dependence below around 130 K. The resistivity curve cannot be described by a thermally activated transport behavior with a single energy gap value across the whole temperature range 10K-310K. However in smaller temperature sub-ranges the data can be reproduced with thermally activated transport with activation gaps from a few to a few tens of meV. These are much smaller than the intrinsic optical gap, and probably related to shallow impurity levels close to the band edges. The inverse Hall resistance 1/eR$_H$ is shown in Fig. 2 (upper inset). The Hall effect is negative, indicating dominant n-type transport, and it is weakly temperature dependent. The room temperature charge carrier density, extracted in a single band model, is around 4x10$^{18}$ cm$^{-3}$. Hall mobility, extracted in a single band description and shown in Fig. 2 (lower inset), is around 40 cm$^2$ V$^{-1}$ s$^{-1}$ at room temperature.

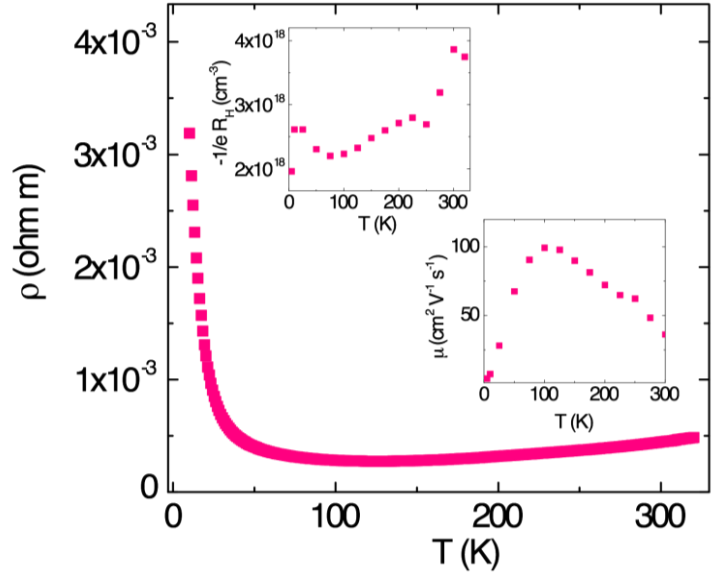

**Fig. 2**: resistivity versus temperature curves measured in the 30 μm thick $SnSe_2$/h-BN flake. Upper inset: Carrier density of the same flakes, extracted from Hall resistance $R_H$. Lower inset: Hall mobility of the same flake.

Fig. 3 (left panel) shows the Seebeck effect measurement of the $SnSe_2$/BN flake. The Seebeck coefficient S is negative and large, comparable to robust thermoelectric materials. At room temperature, the magnitude of S is ≈-400 μV/K. The thermoelectric power factor $S^2\sigma$ is displayed in Fig. 3 (left panel). The room temperature power factor $S^2\sigma \sim 0.35$ mW m$^{-1}$ K$^{-2}$, combined with a low and anisotropic thermal conductivity [7], confirm this compound is promising for thermoelectric applications. This power factor is comparable with literature vaues on bulk-like samples, for example $S^2\sigma \sim 0.15$ mW m$^{-1}$ K$^{-2}$ for a high quality $SnSe_2$ flake [20] and $S^2\sigma \sim 0.8$ mW m$^{-1}$ K$^{-2}$ for highly oriented and dense $SnSe_2$ pellets [22]. However, record-high thermoelectric performances among VdWDs are displayed by few layers thick $MoS_2$ and $WSe_2$ flakes, where S values around 500 μV/K and $S^2\sigma$ values as large as several mW m$^{-1}$ K$^{-2}$ are observed under field effect [9,10,11].

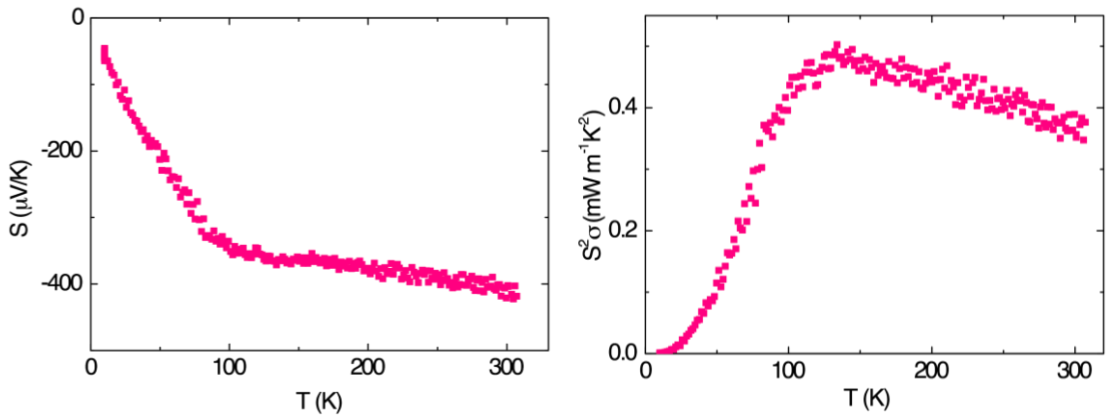

**Fig. 3:** Left: Seebeck coefficient S versus temperatures curves measured in the in the 30 μm thick $SnSe_2$/h-BN flake. Right: Power factor $S^2\sigma$ of the same flake.

In order to get insight into transport mechanisms, we compare the calculated electrical resistivity ρ, thermopower S, power factor $S^2\sigma$, and electron mobility μ values of $SnSe_2$ bulk with experimental results (Figs. 2 and 3). This comparison requires fixing the chemical potential to match the electron carrier density *n* of the meaured sample at 300 K. This yields fair agreement with the measured electric and thermoelectric transport properties, as shown in Fig. 4. At higher temperature reproducing the increasingly metallic resistivity behavior would require other models. The electrical resistivity was estimated considering the temperature *T* and energy $E$ dependences of polar optical and ionized impurity relaxation times (see methods). We find that the $T$ dependence of resistivity in the [120K, 300K] temperature interval is dominated by polar optical scattering: by fitting the experimental electrical resistivity value at 300 K, we estimate a reference polar optical relaxation time $\tau_{po}^{ref}$ ~ 0.38 x 10$^{-14}$ s. At low T ionized impurity scattering dominates: by fitting the experimental electrical resistivity at 20 K we estimated a reference impurity relaxation time $\tau_{imp}^{ref}$~ 8.7 x 10$^{-14}$ s. For comparison the resistivity with acoustic phonon scattering is shown in the same figure, but it gives a decrease of resistivity with T in the [120K, 300K] temperature interval, and is discarded from further fits. Usually, the deformation potentials of acoustic and optical phonon scatterings are dominant at much higher temperature (e. g. for T> 600K). The total relaxation time $\tau$ is given through Matthiessen's rule: $1/\tau = 1/\tau_{po} + 1/\tau_{imp}$. We used the total (energy dependent) relaxation time in the estimation of the resistivity, thermopower, power factor and carrier mobility of bulk SnSe$_2$ (Fig. 4a-d). The latter two are more sensitive to the fit as they involve ratios of n, S, and ρ. Things could be further improved if the temperature dependence of the carrier density $n(T)$ was included, to yield a temperature-dependent $\tau_{po}^{ref}(T)$.

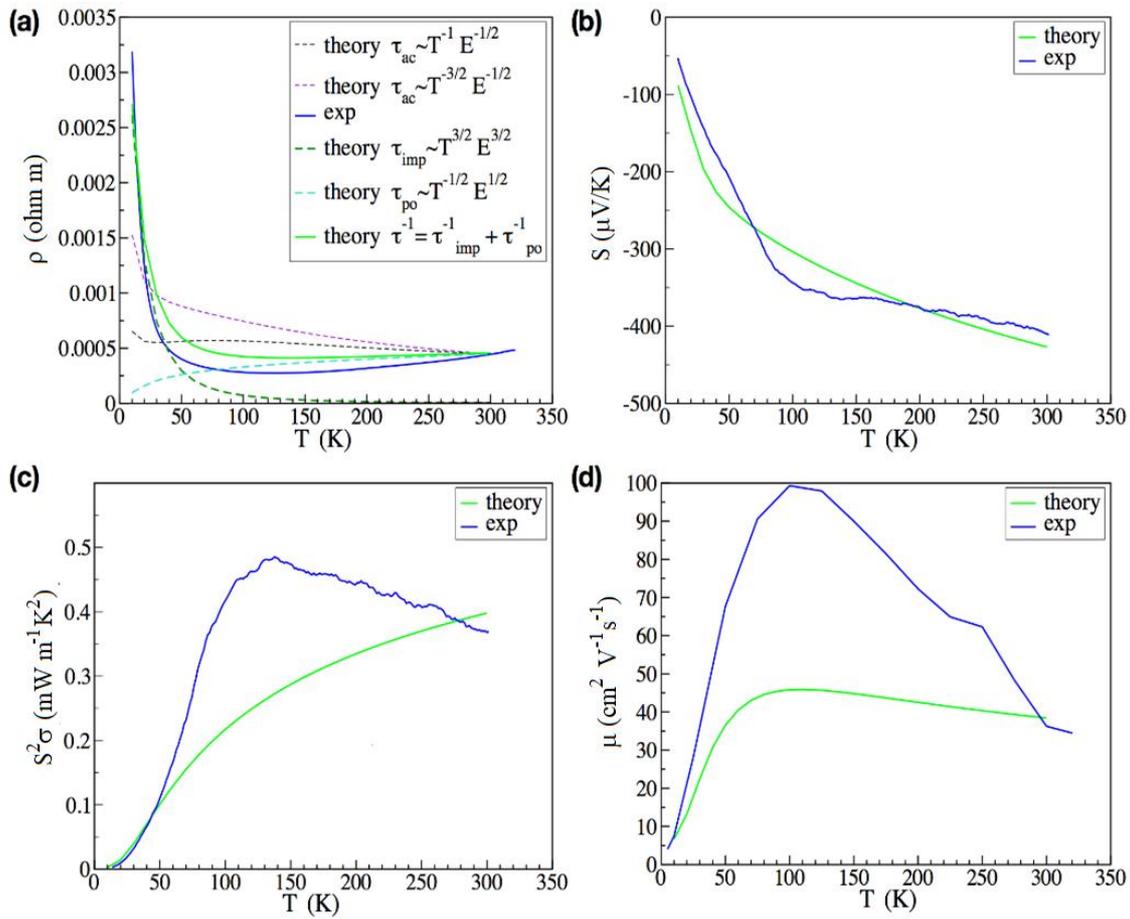

**Fig. 4**: Temperature dependence of thermoelectric properties of SnSe$_2$: (a) resistivity, (b) thermopower S, (c) power factor S$^2$σ, and (d) carrier mobility. The theoretical values are estimated for SnSe$_2$ bulk at an electron carrier density $n$=3.86 x 10$^{18}$ cm$^{-3}$, which matches the experimental carrier density at 300 K. The experimental data correspond to the SnSe$_2$ flake from Figs. 2 and 3.

*b) Tuning of SnSe$_2$ transport properties by field effect*
In order to investigate whether the measured electric and thermoelectric properties can be improved by doping, we carried out reversible electrostatic doping in field effect experiments. To maximize the field effect tuning of the properties, we selected a thinner flake transferred on a substrate with high dielectric permittivity. Fig. 5 (left panel) shows resistivity measurements of a 75 nm SnSe$_2$ flake, exhibiting semiconducting behavior. As for the thicker flake presented in the previous section, it appears that a single transport mechanism cannot describe the whole temperature dependence of resistivity from 10 K to 300 K. The curve can be described by a small polaron hopping mechanism $\rho(T) \propto T exp\left(\frac{E_{SPH}}{k_B T}\right)$ [37] with energy gaps E$_{SPH}$ ≈ 2 meV and 10 meV dominating at low temperatures (T< 120 K) and a thermal activation mechanism $\rho(T) \propto exp\left(\frac{E_{gap}}{k_B T}\right)$ with energy gap E$_{gap}$ ≈ 30 meV dominating at high temperatures (T> 120 K). From Hall effect measurements, we find a n-type carrier density ranging from around 2x10$^{17}$ cm$^{-3}$ at low temperature to 7x10$^{18}$ cm$^{-3}$ at high temperature, presented in Fig. 5 (right panel). Mobility, shown in the inset, is on the order of 1 cm$^2$V$^{-1}$s$^{-1}$. In general, from the comparison of transport properties of the thicker (Fig. 2) and thinner (Fig. 5) flakes, as well as from the characterization of other 15 SnSe$_2$ flakes, it can be said that the effect varying the thickness from several tens of microns to several tens of nanometers is a stronger sensitivity of the transport properties to surface and interface impurities or oxidized superficial layers, so that samples in the tens of nanometers thickness range exhibit higher resistivity and lower mobility.

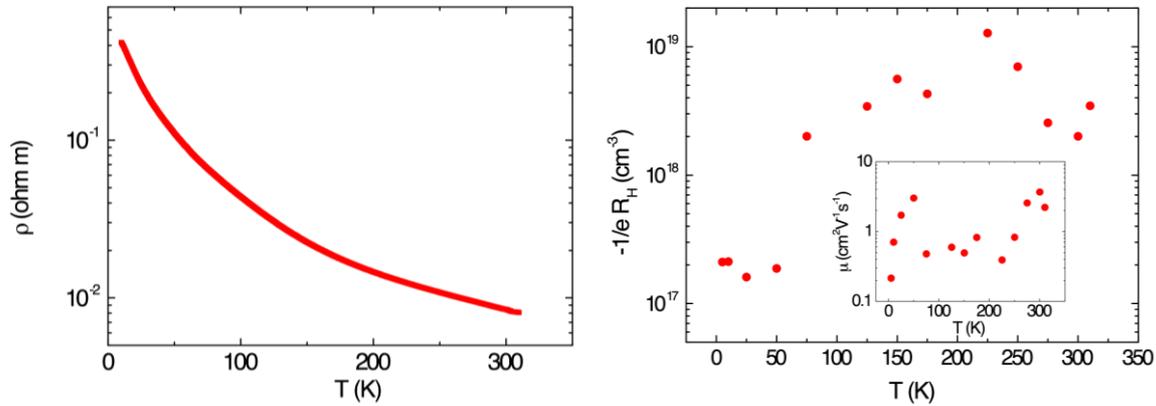

**Fig. 5**: Left: resistivity versus temperature of a 75 nm thick SnSe$_2$ flake. Right: Carrier density (main panel) and mobility (inset) of the same flake, extracted from Hall resistance R$_H$.

The variation of resistance by reversible accumulation and depletion of charge carriers by field effect was measured at different temperatures from 10 K to 200 K. At temperatures higher than 200 K, the experiment could not be performed due to leakage current across the substrate, which is common for SrTiO$_3$ close to room temperature. Resistance versus gate voltage data are displayed in Fig. 6 for the same 75 nm thick SnSe$_2$ flake. With a back gate voltage in the range ± 200 V, the relative variation of resistance ΔR/R=((R(V$_{gate}$=-200V)-R(V$_{gate}$=+200V))/R(V$_{gate}$=0)) is 5.2% at 10 K, 2.5% at 50K, 1.8% at 100K and 0.64% at 200K (see Fig. 6). Hysteresis was observed in different forward and backward sweeps of gate voltage, due to trapping-detrapping of charges at the SrTiO$_3$/SnSe$_2$ interface and possibly also electromigration of charged traps. At all temperatures the measured relative variation of resistance is much smaller than the relative variation of carrier density expected on the basis of a parallel capacitor model, taking into account the carrier concentration measured by Hall effect and the temperature dependence of the SrTiO$_3$ dielectric constant [26]. This discrepancy between expected and measured relative variations is likely due to imperfect adherence of the flake to the substrate during the exfoliation and transfer technique (which can probably be optimized). This leads to the presence of a void dead layer at the SrTiO$_3$/SnSe$_2$ interface and of charged impurity traps at this interface which screen the gate electric field.

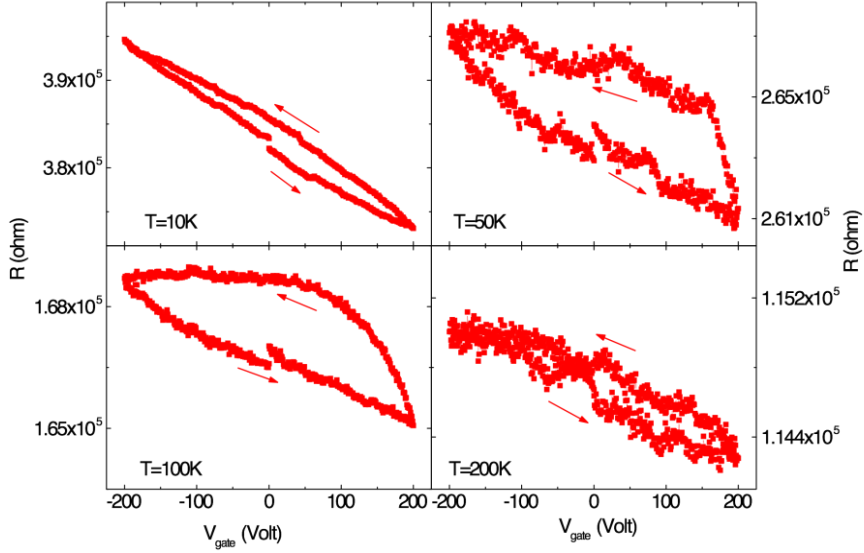

**Fig. 6**: Variation of resistance of the 75 nm thick SnSe$_2$ flake with applied back gate voltage at different temperatures. The arrows indicate the direction of the gate voltage sweep.

In the left panel of Fig. 7, the temperature dependence of the Seebeck coefficient of the SnSe$_2$ flake is shown, exhibiting a maximum as a function of temperature, |S| ≈650 µV/K at 110K, and a room temperature value of |S| ≈10 µV/K. In the right panel, the power factor peaks at S$^2\sigma$~0.014 mW m$^{-1}$ K$^{-2}$ around 125 K. In the thicker flake the carrier density is an order of magnitude smaller, and S almost saturates above 100K (Fig. 3, left).Instead in the thinner flake, with higher carrier density, S decays to a much lower value in approaching room temperature. By fitting the low temperature nearly-linear portions of Seebeck data for the two samples with the Cutler Mott relationship $S = -\left(\frac{3}{2} + \alpha\right)\frac{\pi^2}{3}\frac{k_B^2}{e}T\frac{1}{E_F}$ (where k$_B$ is the Boltzmann constant, $e$ the electron charge, and $\alpha$ the exponent of the power law energy dependence of the scattering time $\tau \propto E^\alpha$), we find Fermi energies E$_F$~ 20 meV for the thin flake and E$_F$~ 29 meV for the thick flake. For a spherical Fermi surface $E_F = \frac{\hbar^2}{2m_{eff}}(3\pi^2 n)^{2/3}$ (where m$_{eff}$ is the effective mass at the Fermi level), hence the difference of the Fermi level values does not fully account for the one order of magnitude difference in carrier concentrations at low temperatures. It is possible that different effective masses at the Fermi level, as well as different scattering mechanisms, as mentioned above, are dominant in the two samples. The thermoelectric power factor S$^2\sigma$ has a maximum at 110 K, with a value 1.1 x 10$^{-2}$ mW m$^{-1}$ K$^{-2}$, while it is 1.2 x 10$^{-5}$ mW m$^{-1}$ K$^{-2}$ at room temperature, and 6.0 x 10$^{-6}$ mW m$^{-1}$ K$^{-2}$ at 10 K. The Goldsmid-Sharp criterium [38] can be used to estimate the band gap E$_g$, (E$_g$ = 2 $e$ |S$_{max}$| T$_{max}$), resulting in a band gap of ~0.15 eV, which is much smaller than the intrinsic SnSe$_2$ band gap of ~1 eV (1.2 eV within B1-WC DFT [12]). This suggests the presence of in-gap states in the thinner SnSe$_2$ flake, which may lead to polaronic carriers similar to what was seen in SnSe$_2$ monolayers on Nb-doped SrTiO$_3$ substrates [39]. Such narrow bandwidth polaronic states can explain the low mobilities of the thinner SnSe$_2$ flake (lower inset of Fig. 2). The non monotonic temperature dependence of S (Fig. 7) as well can be explained in terms of the small band gap: near ~110K, S begins to decrease because of electronic thermal excitation across the small band gap of ~0.15 eV, and bipolar (electron and hole) contributions with opposite signs result in the decrease of S.

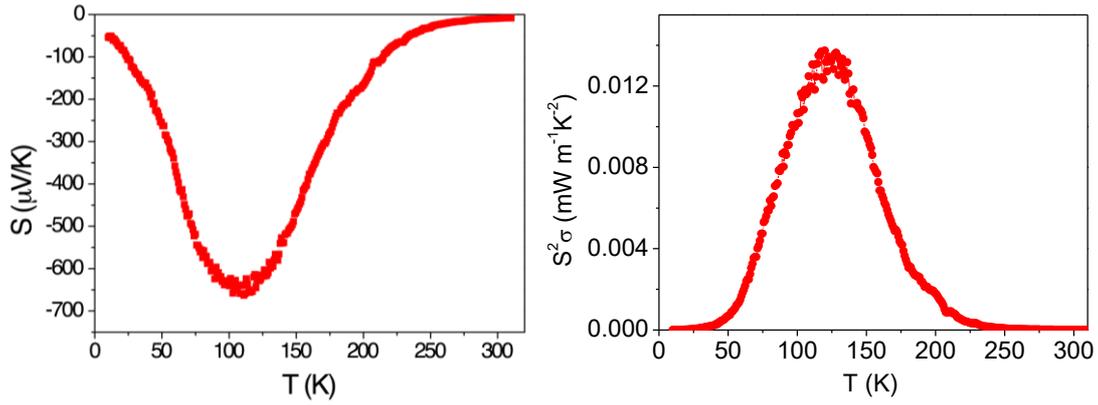

**Fig. 7**: Seebeck coefficient (left panel) and power factor (right panel) of the 75 nm thick SnSe$_2$ flake as a function of temperature.

Fig. 8 shows the variation of S with applied back gate voltage in the range $\pm$ 200 V, at different temperatures. The total relative variation $|\Delta S/S|=((|S|(V_{gate}=-200V)-|S|(V_{gate}=+200V))/|S|(V_{gate}=0))$ is 290% at 10 K, 240% at 50 K, 24% at 100 K and 9% at 200 K (see Fig. 9), much larger than the relative variation of resistance. This discrepancy is not implausible, considering that the variation of S is proportional to the variation of the *energy derivative* of the conductivity, instead of its absolute magnitude. Similar field effect were observed in other semiconducting VdWDs in literature [10,40,41], where variations in S were larger than in conductivity. It is worth noting that at the lower temperatures where the $|\Delta S/S|$ variation is largest the $|\Delta S/S|$ versus $V_{gate}$ curves are asymmetric, with larger variations in the depletion regime (negative $V_{gate}$) than in the accumulation regime (positive $V_{gate}$). This could be a consequence of the Fermi level approaching the conduction band in the accumulation regime, thus sweeping an energy range of higher density of states.

We can estimate the variation of the power factor, to reveal the trend of thermoelectric properties with increasing or decreasing carrier concentration. We find that $S^2\sigma$ moves to higher values with decreasing electron doping, i.e. in the depletion regime, and to lower values in the accumulation regime, with a total variation by field effect that increases with decreasing temperature, also due to the higher dielectric constant of the SrTiO$_3$ substrate. The maximum variation of $S^2\sigma$ is a remarkable 1050 % at 10 K. The reason for the increase of $S^2\sigma$ with decreasing electron density must be ascribed to the corresponding increase of Seebeck coefficient, and the relative change in $S^2\sigma$ is highest at low temperatures, where $S^2\sigma$ is smallest. The relative variations we obtain for the field effect on the Seebeck coefficient, resistance, and power factor are summarized in Fig. 9.

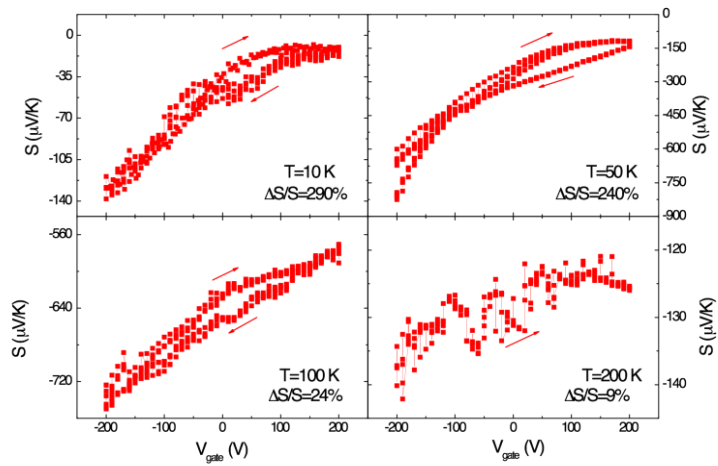

**Fig. 8**: Variation of Seebeck coefficient of the 75 nm thick SnSe$_2$ flake with applied back gate voltage at different temperatures.

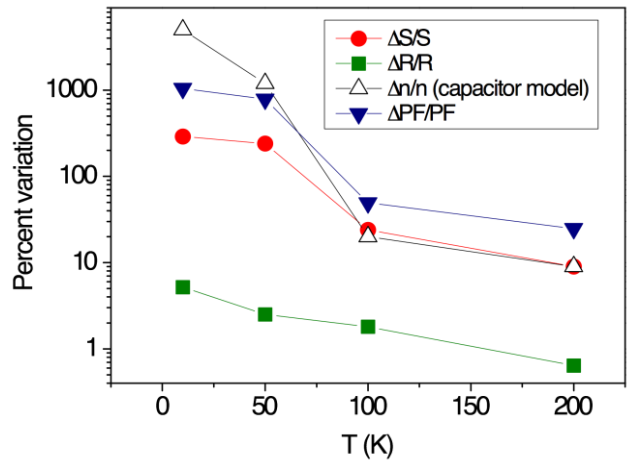

**Fig. 9:** Percent variation of Seebeck coefficient, resistance, and thermoelectric power factor under field effect up to ±200V, for the 75 nm thick SnSe$_2$ flake, compared to the expected percent variation of the charge carrier density, calculated on the basis of a parallel plate capacitor model.

**Conclusions**

In this work, we present a combined experimental and theoretical approach to investigate the electric and thermoelectric properties of SnSe$_2$ flakes. We use the electric field effect in a back gate geometry to cleanly and reversibly tune the electric and thermoelectric properties by band filling and depletion.

SnSe$_2$ exfoliated flakes of micrometric and nanometric thickness exhibit n-type transport and semiconducting temperature dependence of the resistivity. In micrometric thick flakes capped with BN, at room temperature we measure mobility around 40 cm$^2$ V$^{-1}$ s$^{-1}$, a carrier density around 4x10$^{18}$ cm$^{-3}$, a Seebeck coefficient S≈-400 µV/K and thermoelectric power factor S$^2$σ≈0.35 mW m$^{-1}$ K$^{-2}$. Our theoretical calculations of the electric and thermoelectric transport properties are in fair agreement with the measured data. From the calculations, it turns out that the dominant carrier scattering mechanisms are polar optical phonons in the 50-300K temperature interval ($\tau_{po}^{ref}$ of 0.38 x 10$^{-14}$ s at 300K), and ionized impurities below 50 K ($\tau_{imp}^{ref}$ ~ 8.7 x 10$^{-14}$ s at 20 K). For field effect experiments, a flake of nanometric thickness is chosen, exhibiting a Seebeck coefficient that peaks around 100 K with a value |S|≈650 µV/K and a corresponding thermoelectric power factor S$^2$σ≈0.011 mW m$^{-1}$ K$^{-2}$. These field effect experiments indicate that an improvement of thermoelectric properties can be achieved by doping. We find that the power factor increases with depletion of n-type charge carriers. We measure a field effect variation of the Seebeck coefficient up to 290 % at low temperature, and a corresponding variation of the thermoelectric power factor up to 1050 %. The ideal scenario would be to exploit an intermediate flake thickness (in the hundreds of nm), with the favorable starting point of bulk SnSe$_2$, but also a sensitivity to field effect tuning.

**Acknowledgements**

This work was financially supported by the FLAG-ERA JTC2017 project MELODICA "Revealing the potential of transition metal dichalcogenides for thermoelectric applications through nanostructuring and confinement" (http://www.melodica.spin.cnr.it/). DIB acknowledges financial support from a grant of the Romanian National Authority for Scientific Research and Innovation, CCCDI – UEFISCDI, project number COFUND-FLAGERA II-MELoDICA, within PNCDI III. Computational resources were provided by the high performance computational facility of Babes-Bolyai University (MADECIP, POSCCE COD SMIS 48801/1862) cofinanced by the European Regional Development Fund. MJV acknowledges funding from ULiege and the Federation Wallonie Bruxelles through ARC grant DREAMS (G.A. 21/25-11), and a PRACE award granting access to MareNostrum4 at Barcelona Supercomputing Center (BSC), Spain (OptoSpin project id. 2020225411). FC acknowledges the European Union's Horizon 2020 research and innovation programme under the Marie Skłodowska-Curie grant agreement No 892728. IP acknowledges the "Network 4 Energy Sustainable Transition – NEST" project, award number PE0000021, funded under the National Recovery and Resilience Plan (NRRP), Mission 4, Component 2, Investment 1.3 - Call for tender No. 1561 of 11.10.2022 of Italian Ministero dell'Università e della Ricerca (MUR); funded by the European Union – NextGenerationEU.